# Spectral Intensities of Antiprotons and the lifetime of Cosmic Rays in the Galaxy.

R. Cowsik and T Madziwa-Nussinov

Physics Department and McDonnell Center for the Space Sciences

Washington University, St. Louis, MO 63130

*In this paper\* we note that the spectral intensities of antiprotons observed in Galactic cosmic rays in the energy range ~ 1-100 GeV by BESS, PAMELA and AMS† instruments display nearly the same spectral shape as that generated by primary cosmic rays through their interaction with matter in the interstellar medium, without any significant modifications. More importantly, a constant residence time of ~ 2.5 ± 0.7 million years in the Galactic volume, independent of the energy of cosmic rays, matches the observed intensities. A small additional component of secondary antiprotons in the energy below 10 GeV, generated in cocoon-like regions surrounding the cosmic-ray sources, seems to be present. We discuss this result in the context of observations of other secondary components like positrons and Boron‡, and conclude with general remarks about the origins and propagation of cosmic rays.*

---

\* Abstract submitted to the 34th ICRC, Hague, Netherlands, March 19, 2015.
† Amended to include recently released $\bar{p}$ data by AMS.
‡ After we uploaded version 1, we could access the more extensive AMS data on B/C ratio. We show an exclusive fit to this new data and the fits to the antiproton spectrum and ratio implied by the new parameters, which yield equally good fits, as those given in version I.

**Analysis of the spectral intensities of cosmic-ray antiprotons.**

The spectral intensities of antiprotons in cosmic rays provide a crucial diagnostic for understanding the origin and propagation of cosmic rays and complement the information obtained by studying the B/C ratio, the positron fluxes, and the anisotropies in the intensities at higher energies. Even though antiproton fluxes have been measured for several decades, it is only recently that good data, acquired with the PAMELA [1], BESS [2] and AMS [3-6] instruments, have become available over a wide energy range. We have included the $\bar{p}/p$ ratio measured by the AMS instrument along with those from PAMELA and BESS in Fig. 2. In order to bring out some important aspects of the production of cosmic-ray antiprotons, we first show in Fig. 1, the observed spectral intensities of antiprotons, followed by the observed $\bar{p}/p$ ratio in cosmic rays, including the recently released AMS data [5] in Fig. 2.

Our calculation of the spectral intensities of antiprotons follows the simple model that we adopted earlier for the interpretation of the B/C ratio and the spectra of positrons and electrons [7, 8]. It envisages a large number of discrete sources that accelerate the primary cosmic rays and are sprinkled across the Galactic disc. Each of these sources is surrounded by a cocoon-like region where some spallation of cosmic-ray nuclei take place, however without any reacceleration. In this cocoon, the transport of cosmic rays is dependent upon energy, with the diffusion rate increasing with increasing energy, as exemplified by the familiar transport of cosmic rays in the heliosphere. Once cosmic rays leak out into the general interstellar medium, their subsequent transport is independent of the energy of the particles. We characterize the transport in these regions [7, 8] in terms of effective residence times, $\tau_C(E)$ in the cocoon and $\tau_G$ in the Galactic volume:

$$\tau_C(E) \approx \tau_0 T^{(0.01-\zeta \cdot ln T)} \qquad (1)$$

and

$$\tau_G = \text{constant} \qquad (2)$$

with the kinetic energy, T, in GeV per nucleon or per electron or positron as the case may be and $\zeta \approx 0.1$. The values of various parameters in Eq. 1 and Eq. 2 were determined by fitting the B/C ratio in cosmic rays (Fig. 3); the energy dependent part of the B/C ratio was attributed to the spallation in the cocoon and the energy independent fraction to that in the interstellar medium of the Galaxy. The kinematics of the production of B, e$^+$ and $\bar{p}$ in high-energy particle collisions dictate their contributions to the observed spectrum due to interactions in the material inside the cocoon relative to that in the interstellar medium. For example, the B nuclei emerge from the spallation reactions with the same energy per nucleon as their parent nuclei such as C, O etc. In contract, the positrons carry away on average only about 5% of the energy of their parent nucleons. As the cosmic rays leak away from the cocoon progressively more rapidly with increasing energy, very few of them interact at energies needed to generate positrons with energies larger than a couple of GeV. Thus the cocoons inject very few high-energy positrons into the ISM. On the other hand the spallation in the cocoon contributes nearly 75% to the observed B/C ratio at around 1 GeV/nucleon. This contribution becomes progressively smaller at higher energies as their parent nuclei leak away more rapidly from the cocoon experiencing a lesser degree of spallation at higher energies. Since there is a nearly a linear one-to-one correspondence between the energies of B and their parent nuclei, the falling part of the ratio traces the energy dependence of the leakage rate from the cocoon, at energies higher than 1 GeV, where the energy lost due to ionization may be neglected.

The situation with regard to antiprotons is more complex compared with either the B nuclei or the positrons: The threshold for the production of the antiprotons in the proton-proton collisions is 7 $m_p \approx$ 6.6 GeV. Even at threshold, the antiproton emerges in the laboratory frame with an energy of 2 $m_p \approx$ 1.88 GeV. In the high-energy limit, the antiprotons emerge from the interaction with a very broad distribution of kinetic energies in the laboratory frame, ranging from just 100 MeV to almost the entire energy of the incoming proton. The effect of such kinematics is that the dominant contribution at all energies arises through the interactions of cosmic rays with matter in the interstellar medium. In addition, in the low-energy region, below about 10 GeV, there is a noticeable contribution of antiprotons arising from the nuclear interaction of primary cosmic rays in the cocoons surrounding the sources.

Once the cosmic rays leak into the ISM of the Galaxy, their transport does not depend on their energies up to ~ 100 TeV. This transport and the subsequent leakage into the intergalactic space is characterized by an effective residence time $\tau_G$. All the antiprotons observed in cosmic rays at energies above ~ 10 GeV are essentially generated in the ISM in the interactions of the Galactic cosmic rays. At these energies, scaling of their production cross-section ensures that the $\bar{p}/p$ ratio remains constant and any allowable violations are hardly discernable. In this brief note we provide an outline of our calculations and defer a full description to a paper that will be put out in the near future.

For the purposes of this calculation, we have made use of the empirical fits to the inclusive cross-sections for the $\bar{p}$ production in p-p collisions. Besides the many empirical prescriptions available in the past [9], the ones by Mauro et al. [10] and Kappl and Winkler [11] are the most recent and these incorporate the data on antiproton production from the NA49 experiment [12] at CERN. We have adopted the Kappl and Winkler prescription [11] for the cross-sections and these include the contributions arising from the decay of antineutrons and antihyperons to the antiproton flux and incorporate possible violations of scaling at high energies. For the estimation of the contributions of heavier nuclei like He acting as both targets and projectiles, we adopt the calculations of Simon et al. [13] represented by the empirical formula given by Moskalenko and Strong [14]. We assume a mean interstellar flux of cosmic ray protons given by

$$F(E) \approx F_0 E^{-2.7} \text{ cm}^{-2} \text{ s}^{-1} \text{ sr}^{-1} \text{ GeV}^{-1} \qquad (3)$$

at energies higher than 10 GeV as a reasonable fit to a recent compilation of cosmic-ray fluxes [15]. With these assumptions, it is straightforward to estimate the contributions of cosmic-ray interactions with the interstellar gas to the flux of antiprotons. This flux is shown in Fig. 1 as the dashed line, for a mean interstellar density, $n_H \approx 0.5$ and a residence time $\tau_G \approx 2.35$ Myr corresponding to the dashed line in Fig. 3. This contribution of the interstellar production to the antiproton spectrum fits well the observations at high energies and displays a spectral index ~ 2.7, the same as that of the primary cosmic rays.

On the other hand, at energies below ~ 10 GeV, the observed antiproton fluxes are noticeably higher than the estimated values due to production in the interstellar medium. This difference is made up by the contributions to the $\bar{p}$ fluxes generated by cosmic-ray interactions in the material of the cocoons surrounding the sources. As the cosmic rays leak away from the cocoon more rapidly at higher energies, fewer of them will interact at progressively higher energies and the spectrum of the antiprotons generated in the cocoon will be steeper. The grammage in the cocoon relative to that in the interstellar medium is essentially the same as the relative production of boron nuclei in the cocoon with respect to that in the interstellar medium. From Fig. 3, we find that at 1 GeV/nucleon the grammage in the cocoon is 3.7 times larger than in the interstellar medium and falls off with increasing energy as $\tau_c(E)$ given in Eq. 1. The calculation of the $\bar{p}$ fluxes from the cocoons to the observed antiproton spectrum is shown in Fig. 1 as the blue dash-dotted line. The sum of the two contributions first from the cocoon and subsequently from the interactions in the ISM reproduces the observed spectral intensities well. This is shown as the solid purple line in Fig. 1. In order to display the $\bar{p}/p$ ratio, we need a better representation of the spectral intensities of the cosmic-ray protons below ~ 10 GeV than that given in Eq. 3 representing the high energy behavior, as the actual intensities fall below the power-law extrapolation of the empirical fit to lower energies. We use the actual measured intensities [15] to generate Fig. 2 where the theoretical $\bar{p}/p$ ratio is displayed along with the observed ratios [1, 2] including the newly announced data by the AMS group [5]. The good fit to the observations is evident, especially the flat dependence of the ratio at high energies. The observational data are as yet not accurate enough to establish any weak logarithmic increase of cross-sections expected due to scaling violations in some models.

**Discussion and Conclusions.**

The spectral intensities of antiprotons in cosmic rays and their ratio with respect to the intensities of their parents, mainly the cosmic ray protons are well explained as due to a combination of secondary generation in the cocoons surrounding the sources and secondary generation in the ISM. A crucial assumption in providing this explanation is that the leakage lifetime of cosmic rays from the Galaxy is essentially independent of energy from ~ 1 GeV up to ~ 100 TeV, while leakage lifetime from the cocoon decreases with increasing energy. The flatness of the observed $\bar{p}/p$ ratio at energies greater than ~10 GeV provides strong evidence for this energy-independent resident time for cosmic rays in the Galactic volume. In our calculations we have not included any adiabatic losses due to convection or energy gains due to stochastic or other acceleration processes or indeed any process that changes the spectral shape of the antiprotons. In this sense it is a minimal model. We have adopted this model earlier [7, 8] to interpret the B/C ratio (see Fig. 3), the positron spectra and the positron fraction (see Fig. 4) and the bounds on anisotropy of cosmic rays (see Fig. 5). The model described here is generally referred to as the nested leaky-box model (NLB) for cosmic rays [16].

The same energy-independent lifetime, $\tau_G$, in the Galaxy and the energy-dependent lifetime, $\tau_C(E)$, in the cocoon fit all the four sets of data, namely (1) B/C ratio, (2) the positron spectrum and the positron fraction, (3) the antiproton spectrum and the $\bar{p}/p$ ratio, and (4) the bounds on the anisotropy of cosmic rays. Here we note that the observed B/C ratios at energies beyond about 300 GeV still have large error bars. The basic parameter, $\tau_G$, in our model is essentially determined by the observed B/C ratio at these high energies. Keeping in mind that corrections due to the spallation in the atmosphere and due to the possible spill over of the dominant C nuclei into the B region are the main source of background and systematic errors, the experiments like CRN and AMS, which operate outside the Earth's atmosphere, are important for fixing the value of $\tau_G$. Even though the CRN detector operated in the Space-Lab outside the atmosphere, the statistical accuracy is very poor, because of the short time exposure and also has very large systematic uncertainty because of the large correction attributed to the spill over of C nuclei into the B-region. An updated version of CREAM is expected to operate outside the Earth's atmosphere and the resulting measurements will also help in fixing $\tau_G$ more accurately. But to-date the AMS instrument provides the best available B/C ratios at high energies. Accordingly, we have fit exclusively the new AMS data on the B/C ratio to derive the parameters of $\tau_G$ and $\tau_c$. We find $\tau_G \approx 2.35$ Myr for $n_H \approx 0.5$ and $\tau_c \approx \tau_0 \, T^{0.01-0.1 \ln T}$, not significantly different from the earlier fit to the data available prior to AMS publication [7]. These fits to the BC ratio are shown in Fig. 3. In Fig. 1 and Fig. 2, we show the spectrum of antiprotons and the $\bar{p}/p$ ratio calculated with the new set of parameters. The differences with respect to the earlier fits are given in revision 1 are barely noticeable.

On the other hand, the observed B/C ratio at the highest energies seems to deviate from the value of the energy independent $\tau_G$ assumed in our paper; this difference needs to be addressed:

1) The significant deviations commence at ~ 300 GeV/n, which correspond to rigidities of 600 GV and 660 GV for $^{10}$B and $^{11}$B respectively. It is at this rigidity that the AMS-analysis is changed over to MDR mode. Furthermore, with the low value of the B/C ratio, the corrections for the spill-over effects of C into B become more important and less certain, as noted in the context of CRN results [20].

2) However, if after extended observations and analysis, the B/C ratio continues to decrease steeply with energy, it will be a matter of serious concern for the Nested Leaky-Box model.

3) At the current stage, the nested Nested Leaky-Box model gives remarkably good fits to the B/C ratio, the antiproton spectrum, the positron spectrum and the bounds on the anisotropy of cosmic rays, with the same set of very few parameters. This minimal model has to yet to be augmented with astronomical observations and placed more firmly in the astrophysical context. Better measurements of the various cross-sections will help in fixing the parameters of the model more firmly. For now, it provides a very good platform for the understanding of cosmic-ray observations.

4) If we should set aside NLB and take up the view that cosmic-ray lifetime in the Galaxy continues to drop steeply, then we need to find answers to the following questions:

   a. Why do the positrons at energies below ~ 250 GeV, where the radiative energy losses are small, and the antiprotons have spectra almost identical to that generated by cosmic-ray interactions? In both cases, a spectral slope of ~ 2.7 fits remarkably well. Furthermore, in the case of the positrons, the index changes over from ~ 2.7 to ~ 3.7 at E > 300 GeV as expected due to radiative energy losses when $\tau_G$ is taken as a constant at the level needed to generate their intensities.

   b. If we want to argue that there are other astrophysical processes that generate the antiprotons and positrons, then these sources should put out such spectra that when added to the secondaries lead precisely to a smooth total spectrum with a spectral index ~ 2.7, the same as that of the primary cosmic rays.

c. How can we understand the strict bounds on the anisotropy of cosmic rays of $\delta < 10^{-3}$ up to energies of ~ $10^5$ GeV, if we were to allow cosmic rays to stream out of the Galaxy with ever increasing rate with increasing energy?

In summary, we may state that the mysteries of origins and the propagation still remain. At the present state of our understanding the Nested Leaky-Box model provides the best platform to understand the observations.

## Figures

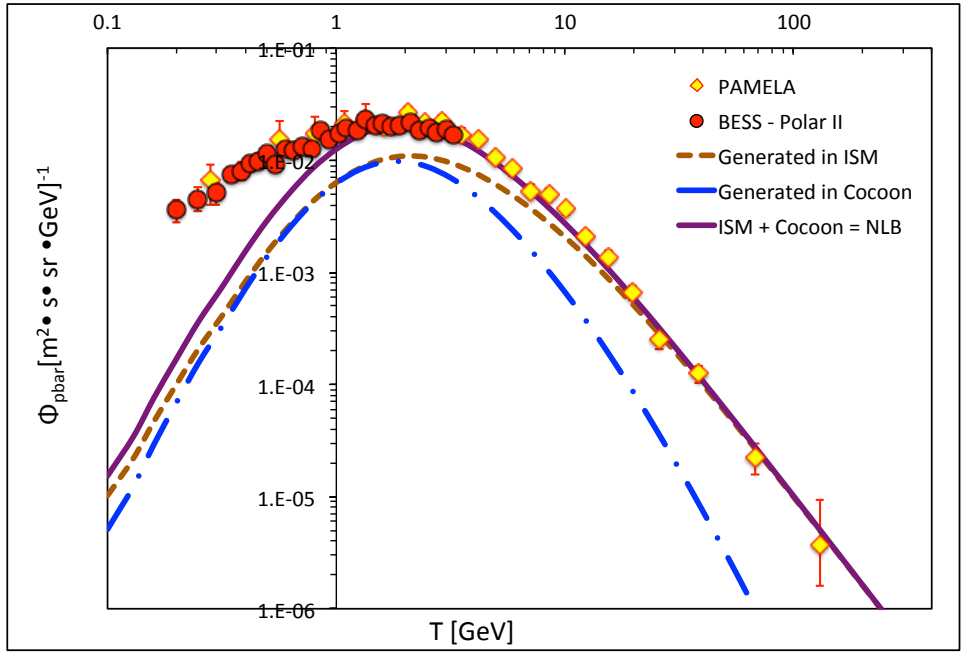

**Fig. 1** - The spectrum of antiprotons observed with the PAMELA and BESS instruments are shown as filled dots [1] and diamonds [2]. In this paper we have interpreted the antiproton spectrum as the sum (solid purple line) of two components: (1) that generated in the ISM (brown dashed line) where the residence time of cosmic rays is independent of their energy and (2) a small component at energies below 10 GeV, with a steep energy dependence at higher energies, generated in a cocoon-like region (blue dash-dotted line) surrounding the sources of primary cosmic rays.

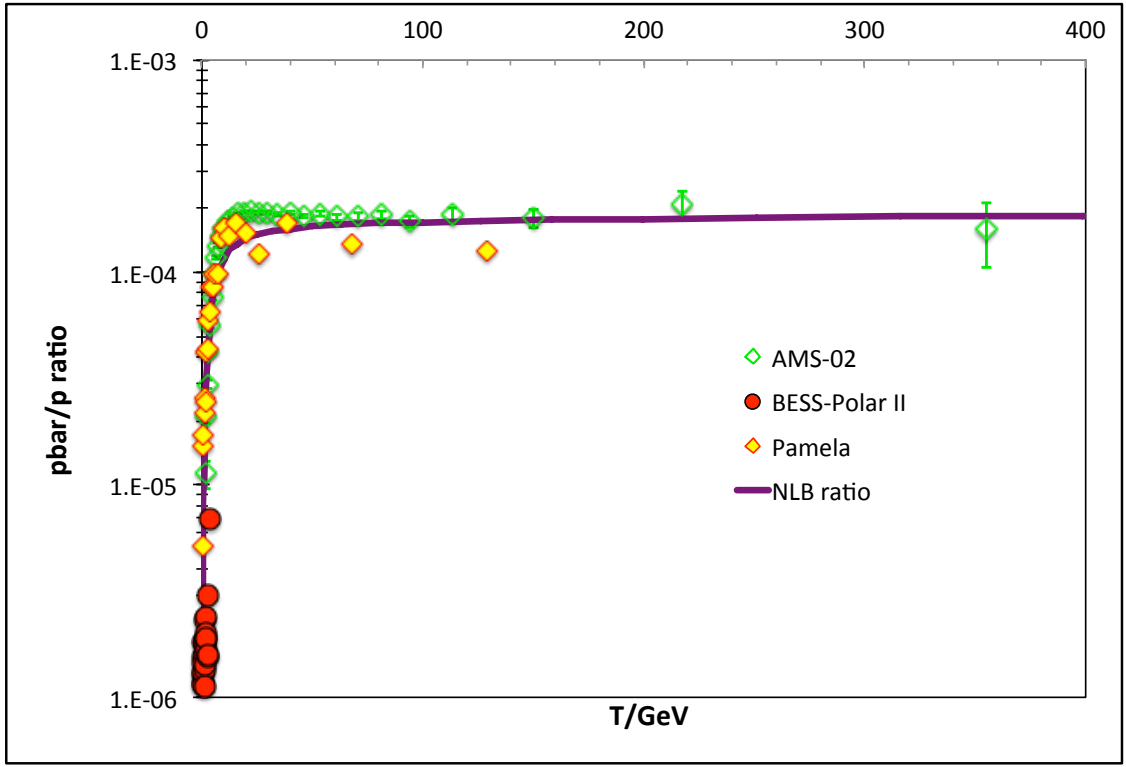

**Fig. 2** – The solid line represents the ratio of the theoretically estimated antiproton flux to the empirical fit to the observed proton flux. The observations from PAMELA, BESS and the AMS-02 instruments [1-5] are shown. The flatness of the $\bar{p}/p$ ratio implies constancy of the leakage lifetime with energy for cosmic rays from the Galaxy.

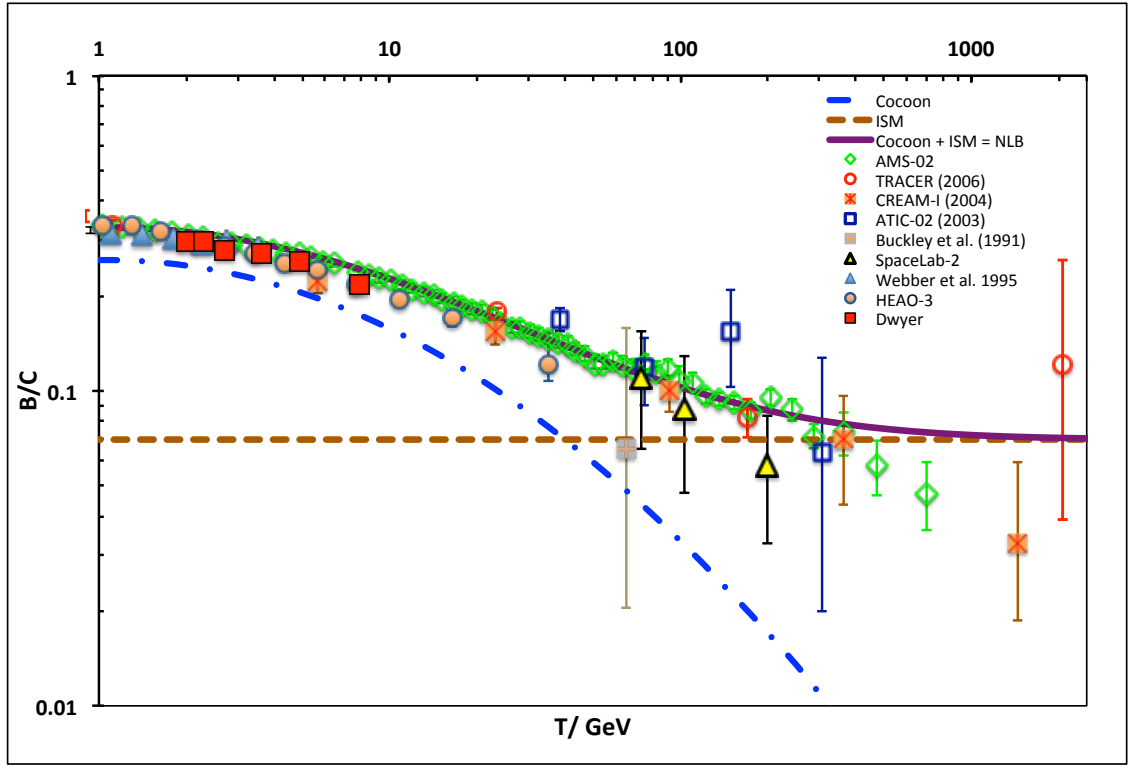

**Fig. 3** – The observed B/C ratio is plotted along with the spectra expected nested leaky-box (NLB) model. The B/C data presented is taken from HEAO-3 [17], Dwyer [18], Chapell and Webber [19], as well as the Tracer [20], Spacelab-2 [21], and CREAM [22], ATIC [27], Buckley [28] and AMS experiments. Two alternate fits are shown, one to the data prior to 2014 (thin lines) [7], and another exclusively to the recent AMS data [6] according to Eq. 1 and Eq. 2 of this paper. These alternate fits have barely discernable effects on the spectrum of antiprotons and $\bar{p}$ to $p$ ratio.

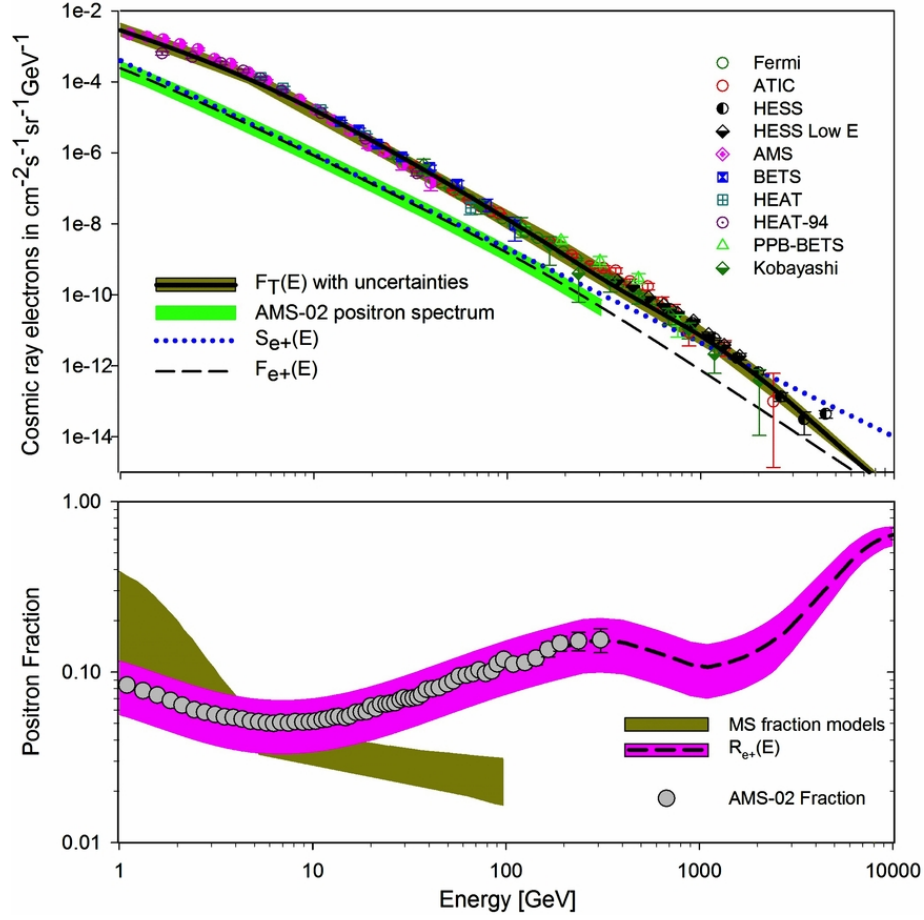

**Fig. 4** – Upper panel: the solid black line represents our fit, $F_T(E)$, to the spectrum of the total electronic component observed in cosmic rays; the band enveloping the dashed and dotted lines show the observed positron spectrum $F_{AMS}$ obtained by multiplying the positron fraction by $F_T(E)$. The dashed line represents the theoretical spectrum, $F_{e+}(E)$, (given in Equation (12) of Cowsik et. al [8]) and the dotted line represents the spectral shape of positrons at production. Lower panel: our predicted positron fraction, $R_{e+}(E) = F_{e+}(E)/F_T(E)$, with uncertainties is shown; the shaded steeply falling region is due to calculations by Moskalenko and Strong using an alternative model for the propagation of cosmic rays. This figure is taken from Cowsik et al. [8].

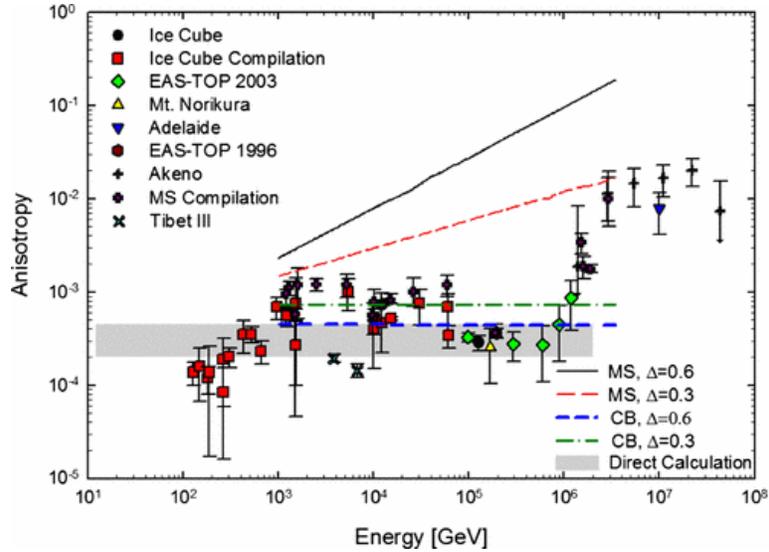

**Fig. 5** - Measurements of the cosmic-ray anisotropy from various compilations [23 - 26]. Also plotted are the predictions from models in M-S [23] and the results from Eq. (7) of Cowsik and Burch [6], which are labeled as CB. The gray region shows the predicted anisotropy from Eq. (8), of Cowsik and Burch [7]. Figures based on references [23-26] as compiled in reference [7].

**Acknowledgements:** The authors wish to thank Professor M. H. Israel for extensive discussions and Professor W. R. Binns for several consultations regarding cosmic rays.